\documentclass{article}
\usepackage{amsmath}
\usepackage{amssymb}
\usepackage{amsfonts}
\usepackage{graphicx}

\begin{document}

\begin{center}
\textbf{Solvable nonlinear systems of }$2$ \textbf{recursions displaying
interesting evolutions}

\bigskip

\textbf{Francesco Calogero}

Dipartimento di Fisica, Universit\`{a} di Roma "La Sapienza", Rome, Italy

Istituto Nazionale di Fisica Nucleare, Sezione di Roma $1$, Rome, Italy

Istituto Nazionale di Alta Matematica, Gruppo Nazionale di Fisica
Matematica, Italy

francesco.calogero@uniroma$1$.it, francesco.calogero@infn.roma$1$.it

\bigskip
\end{center}

\section*{Abstract}
In this paper a class of simple, but nonlinear, systems of recursions
involving $2$ dependent variables $x_{j}\left( n\right) $ is identified,
such that the solutions of their initial-values problems---with arbitrary
initial data $x_{j}\left( 0\right) $---may be explicitly obtained.

\section{\textbf{Introduction}}

In this paper a class of \textit{simple}, but \textit{nonlinear}, systems of 
\textit{recursions} involving $2$ \textit{dependent variables} $x_{j}\left(
n\right) $ is identified, such that the solutions of their \textit{%
initial-values} problems---with \textit{arbitrary} initial data $x_{j}\left(
0\right) $---may be \textit{explicitly} obtained.

\textbf{Notation}: above and hereafter $n$ plays the role of \textit{%
independent variable}, taking \textit{all nonnegative integer} values, $%
n=0,1,2,...$; the $4$ indices $j$, $k$, $s$ and $p$ (the last $2$ only used
in \textbf{Appendix A}) take respectively the $2$ values $j=1,2$, $k=1,2,$ $%
s=1,2$ and $p=1,2$; while the indices $\ell,m,q$ take respectively the $3$
values $\ell=0,1,2$, the $4$ values $m=0,1,2,3$, and the $5$ values $%
q=0,1,2,3,4$. Hereafter I shall generally assume to work in the context of 
\textit{real }numbers, including of course \textit{integers} (as mentioned
just above); occasionally attention shall be restricted to \textit{rational}
numbers; but I shall occasionally need to also use \textit{complex} numbers,
and in that context I shall use the notation $\mathbf{i}$ to denote the 
\textit{imaginary unit} (hence\textit{\ }$\mathbf{i}^{2}=-1$). $\blacksquare$

To arrive at these results, I firstly introduce a system of $2$ \textit{%
linear} recursions satisfied by the $2$ \textit{dependent} variables $%
y_{k}\left( n\right) $: 
\begin{equation}
y_{k}\left( n+1\right) =a_{k0}+a_{k1}y_{1}\left( n\right) +a_{k2}y_{2}\left(
n\right) ~.  \label{SolvLinRec}
\end{equation}%
This system of $2$ \textit{linear} recursions feature $2\cdot 3=6$ \textit{a
priori} \textit{arbitrary} ($n$-independent) parameters $a_{k\ell }$; its 
\textit{initial-values} problem---with \textit{arbitrary} initial data $%
y_{k}\left( 0\right) $---may of course be \textit{explicitly} solved, as
reported in the following \textbf{Section 2}. Then, I introduce the
following \textit{invertible nonlinear transformation} relating the $2$ sets
of dependent variables $x_{j}\left( n\right) $ and $y_{k}\left( n\right) $:%
\begin{equation}
y_{j}\left( n\right) =\frac{b_{j0}+b_{j1}x_{1}\left( n\right)
+b_{j2}x_{2}\left( n\right) }{c_{j0}+c_{j1}x_{1}\left( n\right)
+c_{j2}x_{2}\left( n\right) }~;  \label{NonlinTrans}
\end{equation}%
and, in \textbf{Section 3}, I report its \textit{explicit inversion}. Note
that these relations (\ref{NonlinTrans}) involve $2\cdot 2\cdot 3=12$ 
\textit{a priori arbitrary} ($n$-independent) parameters $b_{j\ell }$ and $%
c_{j\ell }$.

In \textbf{Section 4}, I identify the new \textit{nonlinear} recursions
satisfied by the variables $x_{j}\left( n\right) $ (see (\ref{Recursxj})),
which are \textit{explicitly} solvable because of the way they have been
obtained. These are the main findings obtained in this paper.

Finally, in \textbf{Section 5} I tersely mention the potential \textit{%
applicative} usefulness of the results reported in this paper; results which
I believe to be new, while the techniques exploited to arrive at them are
fairly standard, and have been utilized over time in too many publications
(by myself and by many other authors and co-authors) to justify reporting a
long list of their previous utilizations (since nowadays these publications
are all easily retrievable via \textit{Google}).

This paper is completed by $2$ \textbf{Appendices}. In \textbf{Appendix A}
the \textit{many} formulas are reported which express the $45$ \textit{%
coefficients} of the system of $2$ \textit{nonlinear} recurrences identified
below (see (\ref{Recursxj})) in terms of the $18$ \textit{parameters} which
identify the \textit{solutions} of the \textit{subclass} of this \textit{%
nonlinear} system of $2$ recurrences that is \textit{explicitly solvable};
and let me re-emphasize that the main finding reported in this paper is that
this \textit{subclass} of the system (\ref{Recursxj}), as \textit{explicitly}
identified below, may be \textit{explicitly solved}, as detailed below. In 
\textbf{Appendix B} a subclass of \textit{explicit examples} of this class
of systems, with their \textit{explicit solutions}, are reported; focusing
on cases when the evolution of the system is \textit{isochronous}.

\bigskip

\section{\textbf{A solvable linear system of }$2$\textbf{\ recursions}}

The \textit{linear} character of the recursion (\ref{SolvLinRec}) implies
that its \textit{initial-values} problem is \textit{explicitly solvable}:
indeed by setting 
\begin{subequations}
\label{Soluzyj}
\begin{equation}
y_{j}\left( n\right) =v_{j}(n)+w_{j}
\end{equation}%
with%
\begin{align}  
w_{1}& =\left[ a_{10}\left( 1-a_{22}\right) +a_{20}a_{12}\right] /A~,
\label{w1} \\
w_{2}& =\left[ a_{20}\left( 1-a_{11}\right) +a_{10}a_{21}\right] /A~,
\label{w2} \\
A& =1-a_{11}-a_{22}+a_{11}a_{22}-a_{12}a_{21}~,  \label{A}
\end{align}%
the \textit{new} dependent variables $v_{j}(n)$ satisfy the \textit{%
homogeneous} system of $2$ recursions 
\begin{equation}  
v_{j}\left( n+1\right) =a_{j1}v_{1}\left( n\right) +a_{j2}v_{2}\left(
n\right) ~.  \label{HomoRec}
\end{equation}%
Then $2$, generally independent, solutions $v_{j}^{\left( +\right) }\left(
n\right) $ and $v_{j}^{\left( -\right) }\left( n\right) $ of this system of $%
2$ \textit{homogeneous} recursions are provided by the formulas 
\end{subequations}
\begin{subequations}
\label{SoluzHomo}
\begin{align}
& v_{j}^{\left( \pm \right) }\left( n\right) =V_{j}^{\left( \pm \right)
}\left( u_{\pm }\right) ^{n}~, \\
& V_{2}^{\left( \pm \right) }=\left[ \left( u_{\pm }-a_{11}\right) /a_{12}%
\right] V_{1}^{\left( \pm \right) }~, \\
& u_{\pm }=\left( a_{11}+a_{22}\pm \Delta \right) /2~,  \label{u+-} \\
& \Delta =\sqrt{\left( a_{11}-a_{22}\right) ^{2}+4a_{12}a_{21}}~,
\label{Delta}
\end{align}%
where $V_{1}^{\left( \pm \right) }$ are $2$\ \textit{a priori arbitrary
constants}.

\textbf{Remark 2-1.} Please note that throughout this paper it is assumed
that the number $A$ does \textit{not} vanish, $A\neq0$, see (\ref{A}); and
also that the number $\Delta$ does \textit{not }vanish, see (\ref{Delta})
and (below) the $4$ eqs. (\ref{SolHomoInit}). $\blacksquare$

The general solution of the system of $2$ \textit{linear} recursions (\ref%
{HomoRec}) is then provided by the following formulas: 
\end{subequations}
\begin{subequations}
\label{SolHomoInit}
\begin{align}
v_{1}\left( n\right) & =V^{\left( +\right) }\left( u_{+}\right)
^{n}+V^{\left( -\right) }\left( u_{-}\right) ^{n}~, \\
v_{2}\left( n\right) & =V^{\left( +\right) }\frac{\left( u_{+}-a_{11}\right) 
}{a_{12}}\left( u_{+}\right) ^{n}+V^{\left( -\right) }\frac{\left(
u_{-}-a_{11}\right) }{a_{12}}\left( u_{-}\right) ^{n}~,
\end{align}
where $V^{\left( +\right) }$ and $V^{\left( -\right) }$ are $2$ \textit{%
arbitrary} parameters; and it can be easily checked that the following
assignments of these $2$ parameters,%
\begin{align} 
V^{\left( +\right) } & =\left[ \left( u_{-}-a_{11}\right) v_{1}\left(
0\right) -a_{12}v_{2}\left( 0\right) \right] /\Delta~,  \label{V1+-} \\
V^{\left( -\right) } & =\left[ -\left( u_{+}-a_{11}\right) v_{1}\left(
0\right) +a_{12}v_{2}\left( 0\right) \right] /\Delta~,  \label{V2+--}
\end{align}
imply that these solutions provide the solution of the \textit{initial-values%
} problem corresponding to a generic assignment of the $2$ \textit{initial
data} $v_{j}\left( 0\right) .$Then via the relations (\ref{Soluzyj})---with (%
\ref{SoluzHomo}) and (\ref{SolHomoInit})---one obtains the \textit{explicit}
solution of the \textit{initial-values} problem for the \textit{linear }%
system of $2$ recursions (\ref{SolvLinRec}).

\textbf{Remark 2-2}. The evolution of the solutions of the \textit{linear}
system of $2$ recursions (\ref{SolvLinRec}) is easily ascertained from the
fomulas reported above, but it is nevertheless worthwhile to outline here
some of its main \textit{qualitative} features, which depend crucially on
the values of the $2$ numbers $u_{\pm}$ (see (\ref{u+-})). Clearly if $%
u_{\pm}$ are \textit{both real} (as implied by the condition $\left(
a_{11}-a_{22}\right) ^{2}+4a_{12}a_{21}>0$; see (\ref{u+-}) and (\ref{Delta}%
)), then the \textit{qualitative} behavior of the evolution of these
solutions is determined in a rather \textit{obvious} way by whether the 
\textit{modulus} of these $2$ numbers is \textit{larger} or \textit{smaller}
than $1$, and also by their signs (see (\ref{SolHomoInit})). Marginally less
trivial---and perhaps more interesting---is the analysis of the qualitative
behavior of these solutions if these $2$ \textit{real} numbers $u_{\pm}$\
are \textit{complex-conjugate} of each other, as clearly happens if 
\end{subequations}
\begin{equation}
\left( a_{11}-a_{22}\right) ^{2}+4a_{12}a_{21}<0~,  \label{CondDeltaIm}
\end{equation}
implying that $\Delta$ is \textit{imaginary }(see (\ref{Delta})) hence
requiring of course as \textit{necessary}, but of course \textit{not
sufficient}, condition, that the $2$ parameters $a_{12}\ $and $a_{21}$ have 
\textit{opposite} signs; in which case it is convenient to introduce the%
\textit{\ modulus} $u$ and phase $\varphi$ of $u_{\pm}$: 
\begin{subequations}
\label{periodic}
\begin{equation}
u_{\pm}=u\exp\left( \pm\mathbf{i}\varphi\right) =u\left[ \cos\left(
\varphi\right) \pm\mathbf{i}\sin\left( \varphi\right) \right] ~,  \label{upm}
\end{equation}%
\begin{equation}  
u=\left\vert \sqrt{a_{11}a_{22}-a_{12}a_{21}}\right\vert
~,~~~-a_{12}a_{21}>0~,  \label{u}
\end{equation}%
\begin{equation}  
\varphi=\arctan\left[ \left\vert \Delta\right\vert /\left(
a_{11}+a_{22}\right) \right] ~,  \label{phi}
\end{equation}
see again (\ref{SoluzHomo}) and (\ref{SolHomoInit}). $\blacksquare$

\textbf{Remark 2-3}. The findings reported above (in this \textbf{Section 2}%
) are \textit{elementary} and certainly to be considered as \textit{well
known}, or at least \textit{rather trivial}. Yet I wonder if anybody \textit{%
ever} noticed that they imply that---in the case described in the preceding 
\textbf{Remark 2-2}, which prevails if the $4$ parameters $a_{jk}$ satisfy
the \textit{single inequality} (\ref{CondDeltaIm})---if the $4$ \textit{%
parameters} $a_{jk}$ satisfy \textit{moreover} the following \textit{single
constraint}, 
\end{subequations}
\begin{subequations}
\label{Isoy}
\begin{equation}
a_{11}a_{22}-a_{12}a_{21}=1~,  \label{uEq1}
\end{equation}
implying $u=1$---then \textit{all} the points of coordinates $y_{1}\left(
n\right) ,y_{2}(n)$ yielded by the system of $2$ \textit{linear recursions }(%
\ref{SolvLinRec})\textit{\ }shall, in the $y_{1}-y_{2}$ Cartesian plane, lie
on a \textit{circle}, whose \textit{center} and \textit{radius} are easily
obtained from the results of \textbf{Section 2}; and \textit{moreover} that,
as $n\rightarrow\infty$, these points shall generally \textit{completely
cover} that \textit{circle}; \textit{unless} the $4$ \textit{parameters} $%
a_{jk}$ satisfy $1$ of the following $2$ \textit{second }(not quite
stringent) \textit{simple constraints:} 
\begin{equation}  
a_{11}+a_{22}=\pm2/\sqrt{1+\left\{ \tan\left[ 2\left( N_{1}/N_{2}\right) \pi%
\right] \right\} ^{2}}~,  \label{PeriodicCase}
\end{equation}
with (just above and hereafter) $N_{1}$ and $N_{2}$ being $2$ \textit{%
arbitrary integers} (of course with \textit{no common factor} and $N_{2}>0$%
), implying $\varphi=2\left( N_{1}/N_{2}\right) \pi$; since then---as
clearly implied by (\ref{SolHomoInit}) with (\ref{periodic}) and (\ref{Isoy}%
)---\textit{all solutions} of the corresponding system of $2$ \textit{linear}
recursions (\ref{SolvLinRec}) are \textit{periodic }with the \textit{same
period }$N_{2}$\textit{, } 
\begin{equation}  \label{Isoy}
y_{j}\left( n+N_{2}\right) =y_{j}\left( n\right) ~;
\end{equation}
implying \textit{isochrony} (if the \textit{integer}\ independent variable $%
n\ $may be considered to represent a \textit{ticking time}). $\blacksquare$

\bigskip

\section{\textbf{The inversion of the change of dependent variables }}

In this \textbf{Section 3} the \textit{inversion} of the change of dependent
variables (\ref{NonlinTrans}) is reported, from the $2$ variables $%
y_{j}\left( n\right) $ to the $2$ variables $x_{j}\left( n\right) $, which
is a quite easy task, yielding the following result: 
\end{subequations}
\begin{align}
x_{j}\left( n\right) & =\frac{\beta_{0}^{\left( j\right) }+\beta
_{1}^{\left( j\right) }y_{1}\left( n\right) +\beta_{2}^{\left( j\right)
}y_{2}\left( n\right) +\beta_{3}^{\left( j\right) }y_{1}\left( n\right)
y_{2}\left( n\right) }{\beta_{0}^{(0)}+\beta_{1}^{(0)}y_{1}\left( n\right)
+\beta_{2}^{(0)}y_{2}\left( n\right) +\beta_{3}^{(0)}y_{1}\left( n\right)
y_{2}\left( n\right) }~;  \notag \\
\beta_{0}^{(1)} & =-b_{10}b_{22}+b_{12}b_{20}~,~~\beta_{1}^{\left( 1\right)
}=-b_{20}c_{12}+b_{22}c_{10}~,  \notag \\
\beta_{2}^{\left( 1\right) } & =b_{10}c_{22}-b_{12}c_{20}~,~~\beta
_{3}^{\left( 1\right) }=-c_{10}c_{22}+c_{12}c_{20}~;  \notag \\
\beta_{0}^{\left( 2\right) } & =b_{10}b_{21}-b_{11}b_{20}~,~~\beta
_{1}^{\left( 2\right) }=b_{20}c_{11}-b_{21}c_{10}~,  \notag \\
\beta_{2}^{\left( 2\right) } & =-b_{10}c_{21}+b_{11}c_{20}~,~~\beta
_{3}^{\left( 2\right) }=c_{10}c_{21}-c_{11}c_{20}~;  \notag \\
\beta_{0}^{(0)} &
=b_{11}b_{22}-b_{12}b_{21}~,~~~\beta_{1}^{(0)}=b_{21}c_{12}-b_{22}c_{11}~, 
\notag \\
\beta_{2}^{(0)} &
=-b_{11}c_{22}+b_{12}c_{21}~,~~~\beta_{3}^{(0)}=c_{11}c_{22}-c_{12}c_{21}~.
\label{xjyj}
\end{align}

\bigskip

\section{\textbf{The system of }$2$\textbf{\ nonlinear recursions satisfied
by the }$2$\textbf{\ dependent variables }$x_{j}\left( n\right) $}

The system of $2$ \textit{nonlinear recursions} then satisfied by the $2$
dependent variables $x_{j}\left( n\right) $ reads as follows: 
\begin{subequations}
\label{Recursxj}
\begin{equation}
x_{j}\left( n+1\right) =P_{4}^{(j)}\left[ x_{1}\left( n\right) ,x_{2}\left(
n\right) \right] /P_{4}^{\left( 0\right) }\left[ x_{1}\left( n\right)
,x_{2}\left( n\right) \right] ~,~~j=1,2~,
\end{equation}%
where the $3$ polynomials $P_{4}^{(\ell )}\left( x_{1},x_{2}\right) $ (each
of them of $4$-\textit{th degree}) are defined as follows:%
\begin{align}
& P_{4}^{(\ell )}\left( x_{1},x_{2}\right) =f_{0}^{(\ell
)}+\sum\limits_{k=1}^{2}\left[ {}\right. f_{1k}^{(\ell
)}x_{k}+f_{2k}^{(\ell )}\left( x_{k}\right) ^{2}+f_{3k}^{(\ell )}\left(
x_{k}\right) ^{3}+f_{4k}^{(\ell )}\left( x_{k}\right) ^{4}\left. {}\right]  
\notag \\
& +x_{1}x_{2}\left\{ {}\right. g_{2}^{\left( \ell \right)
}+\sum\limits_{k=1}^{2}\left[ {}\right. g_{3k}^{(\ell )}x_{k}+g_{4k}^{(\ell
)}\left( x_{k}\right) ^{2}\left. {}\right] +g_{4}^{\left( \ell \right)
}x_{1}x_{2}\left. {}\right\} ~,~~\ell =0,1,2~.  \label{P4el}
\end{align}

\textbf{Remark 4-1}. Of course \textit{anyone} of the $3\cdot\left(
1+2\cdot4+1+2\cdot2+1\right) =45$ coefficients $f_{0}^{\left( \ell\right) }$%
, $f_{1k}^{\left( \ell\right) }$, $f_{2k}^{\left( \ell\right) }$, $%
f_{3k}^{\left( \ell\right) }$, $f_{4k}^{\left( \ell\right) }$ and $g_{2\text{%
,}}^{\left( \ell\right) }$ $g_{3k}^{\left( \ell\right) }$, $g_{4k}^{\left(
\ell\right) }$, $g_{4}^{\left( \ell\right) }$ might be replaced (unless it
vanished to begin with) with the \textit{number} $1$, by dividing \textit{all%
} these $45$ coefficients by \textit{that} coefficient; an operation that
clearly leaves \textit{invariant} the system of $2$ \textit{nonlinear}
recursions (\ref{Recursxj}). And \textit{likewise}, in \textit{each} of the $%
2$ relations (\ref{NonlinTrans}), it is clear that $1$ of the $6$ \textit{%
parameters} $b_{j\ell}$ and $c_{j\ell}$ is \textit{not} going to play any 
\textit{significant} role. The, marginally peculiar (but, I hope,
self-explanatory) notation used in the right-hand side of the formula (\ref%
{P4el}) to identify the $15$ coefficients of each of the $3$ polynomials $%
P_{4}^{(\ell)}\left( x_{1},x_{2}\right) $ has the advantage to preserve the
symmetrical role of the $2$ variables $x_{1}$ and $x_{2}$, and is therefore
convenient to write in more compact form some of the formulas reported
below, in \textbf{Appendix A}. $\blacksquare$

\textbf{Remark 4-2}. The formulas displayed above provide the \textit{%
explicit} solution of the \textit{initial-values} problem for the system of $%
2$ \textit{nonlinear} recursions (\ref{Recursxj}): see the expressions (\ref%
{xjyj}) of the $2$ dependent variables $x_{j}(n)$ in terms of the $2$
dependent variables $y_{k}\left( n\right) $, the expressions (\ref{Soluzyj}%
)-(\ref{SolHomoInit}) of the $n$-dependence of the $2$ variables $%
y_{j}\left( n\right) $, and the \textit{explicit} expressions (\ref%
{NonlinTrans}) (with $n=0$) of the $2$ initial values $y_{j}\left( 0\right) $
in terms of the $2$ initial values $x_{k}\left( 0\right) $. And clearly the
simple set of formulas (\ref{xjyj}) imply that the \textit{quantitative} and 
\textit{qualitative} features of the solutions of the system of $2$ \textit{%
linear} recursions described in \textbf{Section 2} are generally transmitted
to the solutions of the system of $2$ \textit{nonlinear} recursions (\ref%
{Recursxj}). $\blacksquare$

The task of expressing the $3(1+2\cdot 4+1+2\cdot 2+1)=45$ coefficients $%
f_{0}^{\left( \ell \right) }$, $f_{1k}^{\left( \ell \right) }$, $%
f_{2k}^{\left( \ell \right) }$, $f_{3k}^{\left( \ell \right) }$, $%
f_{4k}^{\left( \ell \right) }$ and $g_{2\text{,}}^{\left( \ell \right) }$ $%
g_{3k}^{\left( \ell \right) }$, $g_{4k}^{\left( \ell \right) }$, $%
g_{4}^{\left( \ell \right) }$ featured by the right-hand sides of the $2$
recursions (\ref{Recursxj}) in terms of the $3\cdot 2\cdot 3=18$ \textit{%
parameters }$a_{j\ell },~b_{j\ell },~c_{j\ell }$---in terms of which the 
\textit{explicit} solution (\ref{NonlinTrans}) of the \textit{initial-values}
problem for this system of $2$ \textit{nonlinear} recursions (\ref{Recursxj}%
) is provided above---is trivial but tedious (to make sure that I got it
right I used \textbf{Mathematica}; but their extraction from\textbf{\
Mathematica}, and their transfer to the printed form used below, is rather
delicate). The relevant formulas are reported in \textbf{Appendix A}. The
task of inverting those $45$ formulas---and of ascertaining, to the extent
it will be \textit{feasible}, the \textit{constraints} they imply on the $45$
\textit{coefficients }$f_{0}^{\left( \ell \right) }$, $f_{1k}^{\left( \ell
\right) }$, $f_{2k}^{\left( \ell \right) }$, $f_{3k}^{\left( \ell \right) }$%
, $f_{4k}^{\left( \ell \right) }$ and $g_{2\text{,}}^{\left( \ell \right) }$ 
$g_{3k}^{\left( \ell \right) }$, $g_{4k}^{\left( \ell \right) }$, $%
g_{4}^{\left( \ell \right) }$, is left---at least at the moment---to younger
researchers than myself.

\bigskip

\section{\textbf{Envoy}}

In conclusion, let me re-emphasize that the \textit{simple} properties of
the solutions of the \textit{linear} system of $2$ recursions (\ref%
{SolvLinRec}) detailed in \textbf{Section 2}---including the fact that a
subcase of that system displays the remarkable property of \textit{isochrony}%
---are directly transmitted, via the formulas reported in \textbf{Section 3}%
, to the \textit{subclass} of the general system (\ref{Recursxj}) identified
in this paper\textbf{;} as long as those formulas remain valid: namely, the
denominators in their right-hand sides do \textit{not} vanish (see in
particular the right-hand sides of the first entries in the $2$ sets of eqs.
(\ref{Recursxj}) and (\ref{xjyj})); but note that this potential problem is
less likely to emergence in the present context of nonlinear \textit{%
recursions} (when the independent variables takes \textit{only integer}
values) than in that of nonlinear \textit{ODEs} or \textit{PDEs} (when the
independent variable or variables take a \textit{continuum} of \textit{real}
values).

The \textit{mathematical relevance }of the system of $2$ \textit{nonlinear
recursions} (\ref{Recursxj})\textit{---}as its eventual \textit{applicative} 
\textit{relevance}: both by modeling \textit{interesting phenomenologies} in
various scientific contexts (\textit{economy}, \textit{epidemiology,
physics, chemistry, biology,...)}, or as essential background material to
produce \textit{blueprints} necessary to \textit{manufacture devices}
performing certain \textit{specific tasks}---is something the future shall
maybe reveal. Of possible, more immediate, interest are perhaps further
explorations of \textit{subcases} of that system of $2$ \textit{nonlinear}
recursions (\ref{Recursxj}) which emerge if some of the $18$ \textit{a
priori arbitrary parameters} $a_{j\ell},~b_{j\ell},~c_{j\ell}$ are
appropriately \textit{restricted} (a first hint in this direction is the "%
\textit{isochronous}" case, see \textbf{Remark 2-3} above); and also if some
of the $45$ \textit{coefficients }$f_{0}^{\left( \ell\right) }$, $%
f_{1k}^{\left( \ell\right) }$, $f_{2k}^{\left( \ell\right) }$, $%
f_{3k}^{\left( \ell\right) }$, $f_{4k}^{\left( \ell\right) }$ and $g_{2\text{%
,}}^{\left( \ell\right) }$ $g_{3k}^{\left( \ell\right) }$, $g_{4k}^{\left(
\ell\right) }$, $g_{4}^{\left( \ell\right) }$ are assigned special (for
instance \textit{vanishing}) values (of course consistently with the
restrictions on their values implied by their expressions in terms of the $%
18 $ \textit{a priori arbitrary parameters} $a_{j\ell},~b_{j\ell},~c_{j\ell}$%
, as reported in \textbf{Appendix A}). I plan to try and make some progress
in this direction in the future, but with the slowness implied by my very
old age; as well as in the direction of extending the treatment provided in
this paper to the case of more than just $2$ \textit{recursions}, beginning
from the analogous case of $3$ \textit{nonlinearly-coupled} \textit{%
recursions}. While in \textbf{Appendix B} a \textit{single specific example}
(but still involving quite a few free parameters) is reported of the class
of systems (\ref{Recursxj}), together with the (\textit{isochronous}!)
solution of its \textit{initial-values} problem.

\bigskip

\section{\textbf{Appendix A}}

In this \textbf{Appendix A } the $45$ formulas are reported which express
the $3\cdot15=45$ \textit{coefficients} $f_{0}^{\left( \ell\right) }$, $%
f_{1k}^{\left( \ell\right) }$, $f_{2k}^{\left( \ell\right) }$, $%
f_{3k}^{\left( \ell\right) }$, $f_{4k}^{\left( \ell\right) }$ and $g_{2\text{%
,}}^{\left( \ell\right) }$ $g_{3k}^{\left( \ell\right) }$, $g_{4k}^{\left(
\ell\right) }$, $g_{4}^{\left( \ell\right) }$ (here of course $15=9+6$ with $%
9=1+2\cdot4$ and $6=1+2\cdot2+1$: see (\ref{Recursxj})), in terms of the $%
3\cdot2\cdot3=$ $18$ \textit{parameters }$a_{j\ell},~b_{j\ell},~c_{j\ell}$;
which themselves also identify---via the formulas (\ref{xjyj}), (\ref%
{Soluzyj}), (\ref{SoluzHomo}), (\ref{SolHomoInit})---the class of \textit{%
explicit} solutions of the \textit{initial-values} problem (with \textit{%
arbitrary} \textit{initial data }$x_{j}\left( 0\right) $) of the subclass of
the recursions (\ref{Recursxj}) thus identified. Note that below I use 
\textit{coefficients}, \textit{parameters} and other quantities defined
above (for instance, for the $3\cdot4=12$ parameters $\beta _{m}^{\left(
\ell\right) }$, see (\ref{xjyj})), as well as the $4\cdot\left[ \left(
1+4\cdot2\right) +\left( 1+2\cdot2+1\right) \right] =4\left( 9+6\right)
=4\cdot15=60$ new \textit{parameters} $F_{0m}$, $F_{1km} $, $F_{2km}$, $%
F_{3km}$, $F_{4km}~$and $G_{2m}$, $G_{3km}$, $G_{4km}$, $G_{4m}$ introduced
as follows: 
\end{subequations}
\begin{align*}
& f_{0}^{\left( \ell\right) }=\sum_{m=0}^{3}\left[ F_{0m}\beta_{m}^{\left(
\ell\right) }\right] ~,~~~f_{1k}^{\left( \ell\right) }=\sum_{m=0}^{3}\left[
F_{1km}\beta_{m}^{\left( \ell\right) }\right] ~, \\
& f_{2k}^{\left( \ell\right) }=\sum_{m=0}^{3}\left[ F_{2km}\beta
_{m}^{\left( \ell\right) }\right] ~,~~~f_{3k}^{\left( \ell\right)
}=\sum_{m=0}^{3}\left[ F_{3km}\beta_{m}^{\left( \ell\right) }\right] ~, \\
& f_{4k}^{\left( \ell\right) }=\sum_{m=0}^{3}\left[ F_{4km}\beta
_{m}^{\left( \ell\right) }\right] ~,~~~g_{2}^{\left( \ell\right)
}=\sum_{m=0}^{3}\left[ G_{2m}\beta_{m}^{\left( \ell\right) }\right] ~, \\
& g_{3k}^{\left( \ell\right) }=\sum_{m=0}^{3}\left[ G_{3km}\beta
_{m}^{\left( \ell\right) }\right] ~,~~~g_{4k}^{\left( \ell\right)
}=\sum_{m=0}^{3}\left[ G_{4km}\beta_{m}^{\left( \ell\right) }\right] ~, \\
& ~~~g_{4}^{\left( \ell\right) }=\sum_{m=0}^{3}\left[ G_{4m}\beta
_{m}^{\left( \ell\right) }\right] ~,~~~k=1,2,~~~\ell=0,1,2~.
\end{align*}

The expressions of the $60$ \textit{parameters} $F_{0m}$, $F_{1km}$, $%
F_{2km} $, $F_{3km}$, $F_{4km}$ and $G_{2m}$, $G_{3km}$, $G_{4km}$, $G_{4m}$
in terms of the $3\cdot2\cdot3=$ $18$ \textit{parameters }$%
a_{j\ell},~b_{j\ell },~c_{j\ell}$ follow. Note that\ it turns out to be
possible to write these $60$ parameters via (only!?) $28$ separate formulas,
due to the possibility to write always \textit{compactly} the expressions
for the $2$ specific values $1$ and $2$ of the index $m$ (these $2$ values
are occasionally denoted as $s$, $k$ or $p$ below); and also it is possible
to write more compactly the expressions for the $2$ values, $1$ and $2$, of
the index $k$; but, for some of the formulas written below, this requires
the introduction of the following \textit{notational convention}: if in a
formula there appear both the index $k$ and the index $p$, then that formula
holds for $k=1,2$ with $p$ taking the \textit{alternative} assignment (so,
if $k=1$ then $p=2$ while if $k=2$ then $p=1$). 
\begin{subequations}
\begin{equation}
F_{00}=(c_{10}c_{20})^{2}~,  \label{F00}
\end{equation}%
\begin{equation}
F_{0s}=c_{10}c_{20}(a_{s2}b_{20}c_{10}+a_{s1}b_{10}c_{20}+a_{s0}c_{10}c_{20})~,~~s=1,2~,
\label{F0s}
\end{equation}%
\begin{align}
& F_{03}=(a_{12}b_{20}c_{10}+a_{11}b_{10}c_{20}+a_{10}c_{10}c_{20})\cdot 
\notag \\
& (a_{22}b_{20}c_{10}+a_{21}b_{10}c_{20}+a_{20}c_{10}c_{20})~;  \label{F03}
\end{align}

\end{subequations}
\begin{subequations}
\begin{equation}
F_{1k0}=2c_{10}c_{20}(c_{1k}c_{20}+c_{10}c_{2k})~,  \label{FF1k0}
\end{equation}%
\begin{align}
& F_{1ks}=2a_{s0}c_{10}c_{20}(c_{1k}c_{20}+c_{10}c_{2k})+  \notag \\
& a_{s1}c_{20}(b_{1k}c_{10}c_{20}+b_{10}c_{1k}c_{20}+2b_{10}c_{10}c_{2k})+ 
\notag \\
& a_{s2}c_{10}(b_{2k}c_{10}c_{20}+2b_{20}c_{1k}c_{20}+b_{20}c_{10}c_{2k})~,
\label{F1ks}
\end{align}

\begin{align}
& F_{1k3}=c_{20}\left[ {}\right.
a_{10}a_{22}c_{10}(b_{2k}c_{10}+2b_{20}c_{1k})+  \notag \\
& a_{11}a_{22}(b_{1k}b_{20}c_{10}+b_{10}b_{2k}c_{10}+b_{10}b_{20}c_{1k})+ 
\notag \\
& a_{11}(2a_{21}b_{10}b_{1k}+a_{20}b_{1k}c_{10}+a_{20}b_{10}c_{1k})c_{20}+ 
\notag \\
&
a_{10}(a_{21}b_{1k}c_{10}+a_{21}b_{10}c_{1k}+2a_{20}c_{10}c_{1k})c_{20}%
\left. {}\right] +  \notag \\
& (a_{11}b_{10}+a_{10}c_{10})\left[ {}\right.
a_{22}b_{20}c_{10}+2(a_{21}b_{10}+a_{20}c_{10})c_{20}\left. {}\right] c_{2k}+
\notag \\
& a_{12}\left[ {}\right. 2a_{22}b_{20}c_{10}(b_{2k}c_{10}+b_{20}c_{1k})+ 
\notag \\
& c_{10}(a_{21}b_{1k}b_{20}+a_{21}b_{10}b_{2k}+a_{20}b_{2k}c_{10})c_{20}+ 
\notag \\
& b_{20}(a_{21}b_{10}+2a_{20}c_{10})c_{1k}c_{20}+  \notag \\
& b_{20}c_{10}(a_{21}b_{10}+a_{20}c_{10})c_{2k}\left. {}\right] ~;
\label{F1k3}
\end{align}

\end{subequations}
\begin{subequations}
\begin{equation}
F_{2k0}=\left( c_{1k}c_{20}\right) ^{2}+4c_{10}c_{1k}c_{20}c_{2k}+\left(
c_{10}c_{2k}\right) ^{2}~,  \label{F2k0}
\end{equation}%
\begin{align}
&
F_{2ks}=c_{1k}c_{20}(2a_{s2}b_{2k}c_{10}+a_{s2}b_{20}c_{1k}+a_{s1}b_{1k}c_{20}+a_{s0}c_{1k}c_{20})+
\notag \\
& a_{s2}c_{10}(b_{2k}c_{10}+2b_{20}c_{1k})c_{2k}+  \notag \\
& 2(a_{s1}b_{1k}c_{10}+a_{s1}b_{10}c_{1k}+2a_{s0}c_{10}c_{1k})c_{20}c_{2k}+ 
\notag \\
& c_{10}(a_{s1}b_{10}+a_{s0}c_{10})\left( c_{2k}\right) ^{2}~,  \label{F2ks}
\end{align}

\begin{align}
& F_{2k3}=c_{20}\left[ {}\right.
a_{10}a_{22}c_{1k}(2b_{2k}c_{10}+b_{20}c_{1k})+  \notag \\
&
a_{11}a_{22}(b_{1k}b_{2k}c_{10}+b_{1k}b_{20}c_{1k}+b_{10}b_{2k}c_{1k})+(a_{11}b_{1k}+
\notag \\
& a_{10}c_{1k})(a_{21}b_{1k}+a_{20}c_{1k})c_{20}\left. {}\right] +\left[
{}\right. a_{10}a_{22}c_{10}(b_{2k}c_{10}+2b_{20}c_{1k})+  \notag \\
& a_{11}a_{22}(b_{1k}b_{20}c_{10}+b_{10}b_{2k}c_{10}+b_{10}b_{20}c_{1k})+ 
\notag \\
& 2a_{11}(2a_{21}b_{10}b_{1k}+a_{20}b_{1k}c_{10}+a_{20}b_{10}c_{1k})c_{20}+ 
\notag \\
&
2a_{10}(a_{21}b_{1k}c_{10}+a_{21}b_{10}c_{1k}+2a_{20}c_{10}c_{1k})c_{20}%
\left. {}\right] c_{2k}+  \notag \\
& (a_{11}b_{10}+a_{10}c_{10})(a_{21}b_{10}+a_{20}c_{10})\left( c_{2k}\right)
^{2}+  \notag \\
& a_{12}\left\{ {}\right. a_{22}\left[ {}\right. \left( b_{2k}c_{10}\right)
^{2}+4b_{20}b_{2k}c_{10}c_{1k}+\left( b_{20}c_{1k}\right) ^{2}\left. {}%
\right] +  \notag \\
& a_{20}c_{1k}(2b_{2k}c_{10}+b_{20}c_{1k})c_{20}+  \notag \\
& a_{21}(b_{1k}b_{2k}c_{10}+b_{1k}b_{20}c_{1k}+b_{10}b_{2k}c_{1k})c_{20}+ 
\notag \\
& c_{10}(a_{21}b_{1k}b_{20}+a_{21}b_{10}b_{2k}+a_{20}b_{2k}c_{10})c_{2k}+ 
\notag \\
& b_{20}(a_{21}b_{10}+2a_{20}c_{10})c_{1k}c_{2k}\left. {}\right\} ~;
\label{F2k3}
\end{align}

\end{subequations}
\begin{subequations}
\begin{equation}
F_{3k0}=2c_{1k}c_{2k}\left( c_{1k}c_{20}+c_{10}c_{2k}\right) ~,  \label{F3k0}
\end{equation}%
\begin{align}
& F_{3ks}=2a_{s0}c_{1k}c_{2k}\left( c_{1k}c_{20}+c_{10}c_{2k}\right) + 
\notag \\
& a_{s1}c_{2k}\left(
2b_{11}c_{1k}c_{20}+b_{1k}c_{10}c_{2k}+b_{k0}c_{11}c_{21}\right) +  \notag \\
& a_{s2}c_{1k}\left(
b_{2k}c_{1k}c_{20}+2b_{2k}c_{10}c_{2k}+b_{20}c_{1k}c_{2k}\right) ~,~~s=1,2~,
\label{F3ks}
\end{align}%
\begin{align}
& F_{3k3}=a_{22}b_{2k}c_{1k}(a_{11}b_{11}+a_{10}c_{11})c_{20}+  \notag \\
&
a_{12}b_{2k}c_{1k}(2a_{22}b_{2k}c_{10}+2a_{22}b_{20}c_{1k}+a_{21}b_{1k}c_{20}+a_{20}c_{1k}c_{20})+
\notag \\
& a_{12}\left[ {}\right. a_{20}c_{1k}(2b_{21}c_{10}+b_{20}c_{11})+  \notag \\
& a_{21}(b_{1k}b_{21}c_{10}+b_{1k}b_{20}c_{11}+b_{10}b_{2k}c_{11})\left. {} 
\right] c_{2k}+  \notag \\
& \left[ {}\right. a_{10}a_{22}c_{1k}(2b_{21}c_{10}+b_{20}c_{11})+  \notag \\
& a_{11}a_{22}(b_{1k}b_{21}c_{10}+b_{1k}b_{20}c_{11}+b_{10}b_{2k}c_{11})+ 
\notag \\
& 2(a_{11}b_{11}+a_{10}c_{11})(a_{21}b_{1k}+a_{20}c_{1k})c_{20}\left. {} 
\right] c_{2k}+  \notag \\
& \left[ {}\right.
a_{11}(2a_{21}b_{10}b_{1k}+a_{20}b_{1k}c_{10}+a_{20}b_{10}c_{1k})+  \notag \\
& a_{10}(a_{21}b_{1k}c_{10}+a_{21}b_{10}c_{1k}+2a_{20}c_{10}c_{1k})\left. {} 
\right] \left( c_{2k}\right) ^{2}~;  \label{F3k3}
\end{align}

\end{subequations}
\begin{subequations}
\begin{equation}
F_{4k0}=\left( c_{1k}c_{2k}\right) ^{2}  \label{F4k0}
\end{equation}
$~,$%
\begin{equation}
F_{4ks}=c_{1k}c_{2k}\left(
a_{s2}b_{2k}c_{1k}+a_{sk}b_{1k}c_{2k}+a_{s0}c_{1k}c_{2k}\right) ~,
\label{F4ks}
\end{equation}

\begin{align}
& F_{4k3}=\left(
a_{12}b_{2k}c_{1k}+a_{11}b_{1k}c_{2k}+a_{10}c_{1k}c_{2k}\right) \cdot  \notag
\\
& \left( a_{22}b_{2k}c_{1k}+a_{21}b_{1k}c_{2k}+a_{20}c_{1k}c_{2k}\right) ~;
\label{F4k}
\end{align}

\end{subequations}
\begin{subequations}
\begin{equation}
G_{20}=2\left[ {}\right.
c_{10}c_{21}(2c_{12}c_{20}+c_{10}c_{22})+c_{11}c_{20}(c_{12}c_{20}+2c_{10}c_{22})\left. {}%
\right] ~,  \label{G20}
\end{equation}

\begin{align}
&
G_{2s}=a_{s1}c_{20}(b_{12}c_{11}c_{20}+b_{11}c_{12}c_{20}+2b_{12}c_{10}c_{21}+2b_{10}c_{12}c_{21})+
\notag \\
& 2a_{s1}(b_{11}c_{10}c_{20}+b_{10}c_{11}c_{20}+b_{10}c_{10}c_{21})c_{22}+ 
\notag \\
& a_{s2}\left[ {}\right. b_{22}c_{10}(2c_{11}c_{20}+c_{10}c_{21})+  \notag \\
& 2c_{12}(b_{21}c_{10}c_{20}+b_{20}c_{11}c_{20}+b_{20}c_{10}c_{21})+  \notag
\\
& c_{10}(b_{21}c_{10}+2b_{20}c_{11})c_{22}\left. {}\right] +  \notag \\
& 2a_{s0}\left[ (\right.
c_{12}c_{20}(c_{11}c_{20}+2c_{10}c_{21})+c_{10}(2c_{11}c_{20}+c_{10}c_{21})c_{22}\left. {}%
\right] ~,  \label{G2s}
\end{align}

\begin{align}
& G_{23}=a_{12}\left\{ {}\right. 2a_{22}\left[ {}\right.
b_{20}c_{11}(2b_{22}c_{10}+b_{20}c_{12})+  \notag \\
& b_{21}c_{10}(b_{22}c_{10}+2b_{20}c_{12})\left. {}\right] +  \notag \\
& a_{21}\left[ {}\right.
b_{12}(b_{21}c_{10}c_{20}+b_{20}c_{11}c_{20}+b_{20}c_{10}c_{21})+  \notag \\
& b_{11}(b_{22}c_{10}c_{20}+b_{20}c_{12}c_{20}+b_{20}c_{10}c_{22})+  \notag
\\
&
b_{10}(b_{22}c_{11}c_{20}+b_{21}c_{12}c_{20}+b_{22}c_{10}c_{21}+b_{20}c_{12}c_{21}+
\notag \\
& b_{21}c_{10}c_{22}+b_{20}c_{11}c_{22})\left. {}\right] +  \notag \\
& a_{20}\left[ {}\right. b_{22}c_{10}(2c_{11}c_{20}+c_{10}c_{21})+  \notag \\
& b_{21}c_{10}(2c_{12}c_{20}+c_{10}c_{22})+  \notag \\
& 2b_{20}(c_{11}c_{12}c_{20}+c_{10}c_{12}c_{21}+c_{10}c_{11}c_{22})\left. {} 
\right] \left. {}\right\} +  \notag \\
& a11\left\{ {}\right. 2a_{21}\left[ {}\right.
b_{10}c_{21}(2b_{12}c_{20}+b_{10}c_{22})+  \notag \\
& b_{11}c_{20}(b_{12}c_{20}+2b_{10}c_{22})\left. {}\right] +  \notag \\
& a_{22}\left[ {}\right.
b_{12}(b_{21}c_{10}c_{20}+b_{20}c_{11}c_{20}+b_{20}c_{10}c_{21})+  \notag \\
& b_{11}(b_{22}c_{10}c_{20}+b_{20}c_{12}c_{20}+b_{20}c_{10}c_{22})+  \notag
\\
&
b_{10}(b_{22}c_{11}c_{20}+b_{21}c_{12}c_{20}+b_{22}c_{10}c_{21}+b_{20}c_{12}c_{21}+
\notag \\
& b_{21}c_{10}c_{22}+b_{20}c_{11}c_{22})\left. {}\right] +  \notag \\
& a_{20}\left[ {}\right. b_{12}c_{20}(c_{11}c_{20}+2c_{10}c_{21})+  \notag \\
& b_{11}c_{20}(c_{12}c_{20}+2c_{10}c_{22})+  \notag \\
& 2b_{10}(c_{12}c_{20}c_{21}+c_{11}c_{20}c_{22}+c_{10}c_{21}c_{22})\left. {} 
\right] \left. {}\right\} +  \notag \\
& a_{10}\left\{ {}\right. 2a_{20}\left[ {}\right.
c_{10}c_{21}(2c_{12}c_{20}+c_{10}c_{22})+  \notag \\
& c_{11}c_{20}(c_{12}c_{20}+2c_{10}c_{22})\left. {}\right] +  \notag \\
& a_{22}\left[ {}\right. b_{22}c_{10}(2c_{11}c_{20}+c_{10}c_{21})+  \notag \\
& b_{21}c_{10}(2c_{12}c_{20}+c_{10}c_{22})+  \notag \\
& 2b_{20}(c_{11}c_{12}c_{20}+c_{10}c_{12}c_{21}+c_{10}c_{11}c_{22})\left. {} 
\right] +  \notag \\
& a_{21}\left[ {}\right. b_{12}c_{20}(c_{11}c_{20}+2c_{10}c_{21})+  \notag \\
& b_{11}c_{20}(c_{12}c_{20}+2c_{10}c_{22})+  \notag \\
& 2b_{10}(c_{12}c_{20}c_{21}+c_{11}c_{20}c_{22}+c_{10}c_{21}c_{22})\left. {} 
\right] \left. {}\right\} ~;  \label{G23}
\end{align}

\end{subequations}
\begin{subequations}
\begin{equation}
G_{3k0}=2\left[ c_{k0}c_{kp}\left( c_{pk}\right) ^{2}+c_{p0}c_{pp}\left(
c_{kk}\right) ^{2}+2c_{kk}c_{pk}\left( c_{k0}c_{pp}+c_{p0}c_{kp}\right) %
\right] ~,  \label{G3k0}
\end{equation}
\bigskip

\begin{align}
&
G_{3kk}=a_{kk}c_{pk}(2b_{kp}c_{kk}c_{p0}+2b_{kk}c_{kp}c_{p0}+b_{kp}c_{k0}c_{pk}+b_{k0}c_{kp}c_{pk})+
\notag \\
& 2a_{kk}(b_{kk}c_{kk}c_{p0}+b_{kk}c_{k0}c_{pk}+b_{k0}c_{kk}c_{pk})c_{pp}+ 
\notag \\
& a_{kp}\left[ {}\right. b_{pp}c_{kk}(c_{kk}c_{p0}+2c_{k0}c_{pk})+  \notag \\
& 2c_{kp}(b_{pk}c_{kk}c_{p0}+b_{pk}c_{k0}c_{pk}+b_{p0}c_{kk}c_{pk})+  \notag
\\
& c_{kk}(2b_{pk}c_{k0}+b_{p0}c_{kk})c_{pp}\left. {}\right] +  \notag \\
& 2a_{k0}\left[ {}\right.
c_{kp}c_{pk}(2c_{kk}c_{p0}+c_{k0}c_{pk})+c_{kk}(c_{kk}c_{p0}+2c_{k0}c_{pk})c_{pp}\left. {}%
\right] ~,  \label{G3kk}
\end{align}%
\begin{align}
&
G_{3kp}=a_{pk}c_{pk}(2b_{kp}c_{kk}c_{p0}+2b_{kk}c_{kp}c_{p0}+b_{kp}c_{k0}c_{pk}+b_{k0}c_{kp}c_{pk})+
\notag \\
& 2a_{pk}(b_{kk}c_{kk}c_{p0}+b_{kk}c_{k0}c_{pk}+b_{k0}c_{kk}c_{pk})c_{pp}+ 
\notag \\
& a_{pp}\left[ {}\right. b_{pp}c_{kk}(c_{kk}c_{p0}+2c_{k0}c_{pk})+  \notag \\
& 2c_{kp}(b_{pk}c_{kk}c_{p0}+b_{pk}c_{k0}c_{pk}+b_{p0}c_{kk}c_{pk})+  \notag
\\
& c_{kk}(2b_{pk}c_{k0}+b_{p0}c_{kk})c_{pp}\left. {}\right] +  \notag \\
& 2a_{p0}\left[ (\right.
c_{kp}c_{pk}(2c_{kk}c_{p0}+c_{k0}c_{pk})+c_{kk}(c_{kk}c_{p0}+2c_{k0}c_{pk})c_{pp}\left. {}%
\right] ~,  \label{G3kp}
\end{align}%
\begin{align}
& G_{3k3}=a_{k0}\left\{ {}\right.
a_{pk}c_{pk}(2b_{kp}c_{kk}c_{p0}+2b_{kk}c_{kp}c_{p0}+b_{kp}c_{k0}c_{pk}+b_{k0}c_{kp}c_{pk})+
\notag \\
& 2a_{pk}(b_{kk}c_{kk}c_{p0}+b_{kk}c_{k0}c_{pk}+b_{k0}c_{kk}c_{pk})c_{pp}+ 
\notag \\
& a_{pp}\left[ {}\right. b_{pp}c_{kk}(c_{kk}c_{p0}+2c_{k0}c_{pk})+  \notag \\
&
2c_{kp}(b_{pk}c_{kk}c_{p0}+b_{pk}c_{k0}c_{pk}+b_{p0}c_{kk}c_{pk})+c_{kk}(2b_{pk}c_{k0}+b_{p0}c_{kk})c_{pp}\left. {} 
\right] +  \notag \\
& 2a_{p0}\left[ {}\right.
c_{kp}c_{pk}(2c_{kk}c_{p0}+c_{k0}c_{pk})+c_{kk}(c_{kk}c_{p0}+2c_{k0}c_{pk})c_{pp}\left. {}%
\right] \left. {}\right\} +  \notag \\
& a_{kk}\left\{ \left[ {}\right. \right.
a_{p0}c_{pk}(2b_{kp}c_{kk}c_{p0}+2b_{kk}c_{kp}c_{p0}+b_{kp}c_{k0}c_{pk}+b_{k0}c_{kp}c_{pk})+
\notag \\
& 2a_{p0}(b_{kk}c_{kk}c_{p0}+b_{kk}c_{k0}c_{pk}+b_{k0}c_{kk}c_{pk})c_{pp}+ 
\notag \\
& 2a_{pk}\left[ {}\right.
b_{kp}c_{pk}(2b_{kk}c_{p0}+b_{k0}c_{pk})+b_{kk}(b_{kk}c_{p0}+2b_{k0}c_{pk})c_{pp}\left. {}%
\right] +  \notag \\
& a_{pp}\left[ {}\right.
b_{kp}(b_{pk}c_{kk}c_{p0}+b_{pk}c_{k0}c_{pk}+b_{p0}c_{kk}c_{pk})+  \notag \\
& b_{kk}(b_{pp}c_{kk}c_{p0}+b_{pk}c_{kp}c_{p0}+b_{pp}c_{k0}c_{pk}+  \notag \\
& b_{p0}c_{kp}c_{pk}+b_{pk}c_{k0}c_{pp}+b_{p0}c_{kk}c_{pp})+  \notag \\
& b_{k0}(b_{pp}c_{kk}c_{pk}+b_{pk}c_{kp}c_{pk}+b_{pk}c_{kk}c_{pp})\left. {} 
\right] \left. {}\right\} +  \notag \\
& a_{kp}\left\{ {}\right. 2a_{pp}\left[ {}\right.
b_{pp}c_{kk}(2b_{pk}c_{k0}+b_{p0}c_{kk})+b_{pk}(b_{pk}c_{k0}+2b_{p0}c_{kk})c_{kp}\left. {}%
\right] +  \notag \\
& a_{pk}\left[ {}\right.
b_{kp}(b_{pk}c_{kk}c_{p0}+b_{pk}c_{k0}c_{pk}+b_{p0}c_{kk}c_{pk})+  \notag \\
& b_{kk}(b_{pp}c_{kk}c_{p0}+b_{pk}c_{kp}c_{p0}+b_{pp}c_{k0}c_{pk}+  \notag \\
& b_{p0}c_{kp}c_{pk}+b_{pk}c_{k0}c_{pp}+b_{p0}c_{kk}c_{pp})+  \notag \\
& b_{k0}(b_{pp}c_{kk}c_{pk}+b_{pk}c_{kp}c_{pk}+b_{pk}c_{kk}c_{pp})\left. {} 
\right] +  \notag \\
& a_{p0}\left[ {}\right.
b_{pp}c_{kk}(c_{kk}c_{p0}+2c_{k0}c_{pk})+b_{p0}c_{kk}(2c_{kp}c_{pk}+c_{kk}c_{pp})+
\notag \\
& 2b_{pk}(c_{kk}c_{kp}c_{p0}+c_{k0}c_{kp}c_{pk}+c_{k0}c_{kk}c_{pp})\left. {} 
\right] \left. {}\right\} ~;  \label{G3k3}
\end{align}

\end{subequations}
\begin{subequations}
\begin{equation}
G_{4k0}=2c_{1k}c_{2k}\left( c_{12}c_{2k}+c_{11}c_{22}\right) ~,  \label{G4k0}
\end{equation}%
\begin{align}
& G_{4ks}=2a_{s0}c_{11}c_{2k}\left( c_{11}c_{22}+c_{12}c_{21}\right) + 
\notag \\
& a_{s1}c_{21}\left(
b_{11}c_{12}c_{2k}+b_{12}c_{11}c_{2k}+2b_{1k}c_{11}c_{22}\right) +  \notag \\
& a_{s2}c_{11}\left(
b_{21}c_{1k}c_{22}+b_{22}c_{1k}c_{21}+2b_{2k}c_{12}c_{21}\right) ~,~~s=1,2
\label{G4ks}
\end{align}

\begin{align}
& G_{4k3}=c_{pk}\left[ {}\right.
a_{k0}a_{pp}c_{kk}(b_{pp}c_{kk}+2b_{pk}c_{kp})+  \notag \\
& a_{kk}a_{pp}(b_{kp}b_{pk}c_{kk}+b_{kk}b_{pp}c_{kk}+b_{kk}b_{pk}c_{kp})+ 
\notag \\
& a_{kk}(2a_{pk}b_{kk}b_{kp}+a_{p0}b_{kp}c_{kk}+a_{p0}b_{kk}c_{kp})c_{pk}+ 
\notag \\
&
a_{k0}(a_{pk}b_{kp}c_{kk}+a_{pk}b_{kk}c_{kp}+2a_{p0}c_{kk}c_{kp})c_{pk}%
\left. {}\right] +  \notag \\
& (a_{kk}b_{kk}+a_{k0}c_{kk})\left[ {}\right.
a_{pp}b_{pk}c_{kk}+2(a_{pk}b_{kk}+a_{p0}c_{kk})c_{pk}\left. {}\right] c_{pp}+
\notag \\
& a_{kp}\left[ {}\right. 2a_{pp}b_{pk}c_{kk}(b_{pp}c_{kk}+b_{pk}c_{kp})+ 
\notag \\
& c_{kk}(a_{pk}b_{kp}b_{pk}+a_{pk}b_{kk}b_{pp}+a_{p0}b_{pp}c_{kk})c_{pk}+ 
\notag \\
&
b_{pk}(a_{pk}b_{kk}+2a_{p0}c_{kk})c_{kp}c_{pk}+b_{pk}c_{kk}(a_{pk}b_{kk}+a_{p0}c_{kk})c_{pp}\left. ) 
\right] ~;  \label{G4k3}
\end{align}%
\end{subequations}
\begin{subequations}
\begin{equation}
G_{40}=\left( c_{11}c_{22}\right) ^{2}+\left( c_{12}c_{21}\right)
^{2}+4c_{11}c_{12}c_{21}c_{22}~,  \label{G40}
\end{equation}

\begin{align}
&
G_{4k}=c_{12}c_{21}(2a_{k2}b_{22}c_{11}+a_{k2}b_{21}c_{12}+a_{k1}b_{12}c_{21}+a_{k0}c_{12}c_{21})+
\notag \\
& a_{k2}c_{11}(b_{22}c_{11}+2b_{21}c_{12})c_{22}+  \notag \\
& 2(a_{k1}b_{12}c_{11}+a_{k1}b_{11}c_{12}+2a_{k0}c_{11}c_{12})c_{21}c_{22}+ 
\notag \\
& c_{11}(a_{k1}b_{11}+a_{k0}c_{11})\left( c_{22}\right) ^{2}~,  \label{G4k}
\end{align}

\begin{align}
& G_{43}=c_{21}\left[ {}\right.
a_{10}a_{22}c_{12}(2b_{22}c_{11}+b_{21}c_{12})+  \notag \\
& a_{11}a_{22}(b_{12}b_{22}c_{11}+b_{12}b_{21}c_{12}+b_{11}b_{22}c_{12})+ 
\notag \\
& (a_{11}b_{12}+a_{10}c_{12})(a_{21}b_{12}+a_{20}c_{12})c_{21}\left. {} 
\right] +  \notag \\
& \left[ {}\right. a_{10}a_{22}c_{11}(b_{22}c_{11}+2b_{21}c_{12})+  \notag \\
& a_{11}a_{22}(b_{12}b_{21}c_{11}+b_{11}b_{22}c_{11}+b_{11}b_{21}c_{12})+ 
\notag \\
& 2a_{11}(2a_{21}b_{11}b_{12}+a_{20}b_{12}c_{11}+a_{20}b_{11}c_{12})c_{21}+ 
\notag \\
&
2a_{10}(a_{21}b_{12}c_{11}+a_{21}b_{11}c_{12}+2a_{20}c_{11}c_{12})c_{21}%
\left. {}\right] c_{22}+  \notag \\
& (a_{11}b_{11}+a_{10}c_{11})(a_{21}b_{11}+a_{20}c_{11})\left( c_{22}\right)
^{2}+  \notag \\
& a_{12}\left\{ {}\right. a_{22}\left[ {}\right. \left( b_{22}c_{11}\right)
^{2}+4b_{21}b_{22}c_{11}c_{12}+\left( b_{21}c_{12}\right) ^{2}\left. {}%
\right] +  \notag \\
& a_{20}c_{12}(2b_{22}c_{11}+b_{21}c_{12})c_{21}+  \notag \\
& a_{21}(b_{12}b_{22}c_{11}+b_{12}b_{21}c_{12}+b_{11}b_{22}c_{12})c_{21}+ 
\notag \\
& c_{11}(a_{21}b_{12}b_{21}+a_{21}b_{11}b_{22}+a_{20}b_{22}c_{11})c_{22}+ 
\notag \\
& b_{21}(a_{21}b_{11}+2a_{20}c_{11})c_{12}c_{22}\left. {}\right\} ~.
\label{G43}
\end{align}

\bigskip

\section{\textbf{Appendix B}}

In this \textbf{Appendix B }a specific example---in fact a rather large
class of such examples---is described, of the system of $2$ \textit{%
nonlinearly-coupled} recursions (\ref{Recursxj}), which displays a very
simple (\textit{isochronous}!) evolution. It is characterized by the
following $2$ specific restrictions on the $4$ parameters $a_{jk}$: 
\end{subequations}
\begin{subequations}
\begin{equation}
a_{12}a_{21}=a_{11}a_{22}-1~,~~~a_{11}+a_{22}=\pm1~.  \label{2IsoConds}
\end{equation}
It is indeed easily seen from the results of \textbf{Section 2} that \textit{%
any generic} assignment of the $4$ parameters $a_{jk}$ satisfying these $2$
restrictions---implied by the requirement that the $2$ parameters $u$ and $%
\varphi$ (see (\ref{periodic})) take the values 
\begin{equation}
u=1~,~~~\varphi=4\pi/3=2\pi N_{1}/N_{2}
\end{equation}
(hence $N_{1}/N_{2}=2/3$; implying $\left[ \tan\left( \varphi\right) \right]
^{2}=3$)---entails that the corresponding evolution of the system of $2$ 
\textit{linear }recursions (\ref{SolvLinRec}) is \textit{isochronous}, 
\textit{all} its solutions---for \textit{generic initial data} $y_{j}\left(
0\right) $---being \textit{periodic} with period $3$: 
\end{subequations}
\begin{subequations}
\begin{equation}
y_{j}\left( n+3\right) =y_{j}\left( n\right) ~.
\end{equation}

Hence, due to the findings reported in this paper, it may also be asserted
that the corresponding system of $2$ \textit{nonlinear} recursions (\ref%
{Recursxj}) is as well \textit{isochronous}, \textit{all} its
solutions---for \textit{generic initial data }$x_{j}\left( 0\right) $%
---being as well \textit{periodic} with period $3$:%
\begin{equation}
x_{j}\left( n+3\right) =x_{j}\left( n\right) ~;  \label{xniso}
\end{equation}
provided of course that the $45$ coefficients $f_{0}^{\left( \ell\right) }$, 
$f_{1k}^{\left( \ell\right) }$, $f_{2k}^{\left( \ell\right) }$, $%
f_{3k}^{\left( \ell\right) }$, $f_{4k}^{\left( \ell\right) }$ and $g_{2\text{%
,}}^{\left( \ell\right) }$ $g_{3k}^{\left( \ell\right) }$, $g_{4k}^{\left(
\ell\right) }$, $g_{4}^{\left( \ell\right) }$ which define the $3$ $4$%
th-degree polynomials $P_{4}^{\left( \ell\right) }\left( x_{1}x_{2}\right) $
which appear in the right-hand-sides of its $2$ \textit{nonlinear}
recursions (\ref{Recursxj}), be expressed, in terms of the $18$ parameters $%
a_{j\ell},~b_{j\ell},~c_{j\ell}$, by the \textit{explicit} formulas reported
in \textbf{Appendix A}, for \textit{any generic assignment} of these $18$
parameters such that the $4$ parameters $a_{jk}$ do satisfy the $2$ \textit{%
simple restrictions} (\ref{2IsoConds}).

For instance (taking the \textit{positive} sign in the right-hand side of
the second of the $2$ eqs. (\ref{2IsoConds})), it is indeed easily seen that
the \textit{solution} of the \textit{initial-values} problem of the \textit{%
homogenous} system of $2$ \textit{linear recursions} (\ref{HomoRec}) reads
then as follows: 
\end{subequations}
\begin{subequations}
\label{vjnAppB}
\begin{align}
& v_{1}\left( n\right) =v_{1}\left( 0\right) \cos\left( 2\pi n/3\right) + 
\notag \\
& \left[ \frac{\left( 1+2a_{11}\right) v_{1}\left( 0\right) +2v_{2}\left(
0\right) }{\sqrt{3}}\right] \sin\left( 2\pi n/3\right) ~, \\
& v_{2}\left( n\right) =v_{2}\left( 0\right) \cos\left( 2\pi n/3\right) - 
\notag \\
& \left\{ \frac{2\left[ 1+a_{11}+\left( a_{11}\right) ^{2}\right]
v_{1}\left( 0\right) +\left( 1+a_{11}\right) a_{12}v_{2}\left( 0\right) }{%
\sqrt{3}~a_{12}}\right\} \sin\left( 2\pi n/3\right) ~;
\end{align}
and it is plain that this solution is \textit{isochronous}, for \textit{any
generic} assignment of the \textit{initial data }$v_{1}\left( 0\right) $ and 
$v_{2}\left( 0\right) $, in the discrete "ticking time" variable $n$, with
period $3$, for any reasonable assignment of the $2$ parameters $a_{11}$ and 
$a_{12}$; while the $2$ parameters $a_{21}$ and $a_{22}$ are given as
follows, in terms of $a_{11}$ and $a_{12}$ (as implied by the $2$ \textit{%
simple} eqs. (\ref{2IsoConds})):%
\begin{equation}  
a_{22}=1-a_{11}~,~~~a_{21}=\left( a_{11}a_{22}-1\right) /a_{12}~.
\label{a22a21}
\end{equation}
And clearly this remarkable property of \textit{isochrony} is transmitted to
the corresponding solution of the system of $2$ \textit{nonlinear}
recursions (\ref{Recursxj}), provided the $45$ coefficients---which identify
the $3$ polynomials, each of $4$-th degree in the $2$ variables $x_{j}\left(
n\right) $, which appear in the right-hand sides of those formulas---are
given by the formulas reported in \textbf{Appendix A}, in terms of the $18$
parameters $a_{j\ell},~b_{j\ell},~c_{j\ell}$; which may themselves be 
\textit{arbitrarily assigned}, except for the $2$ \textit{simple conditions}
(\ref{2IsoConds}) (or, equivalently, (\ref{a22a21})).

It is left to the interested reader to display all the corresponding
relevant formulas, in particular the \textit{explicit} solution $x_{j}(n)$
of the system of $2$ \textit{nonlinear recursions} (\ref{Recursxj}) in terms
of $2$ \textit{generic initial data }$x_{k}(0)$---via firstly using the
formulas (\ref{NonlinTrans}) at $n=0$ to get $y_{j}\left( 0\right) $ in
terms of $x_{k}\left( 0\right) $, next getting $y_{j}\left( n\right) $ by
the relevant formulas of \textbf{Section 2} in terms of $y_{k}\left(
0\right) $ (see (\ref{Soluzyj}), (\ref{SoluzHomo}), (\ref{SolHomoInit}), (%
\ref{periodic}) and (\ref{Isoy})), and finally via the formulas (\ref{xjyj})
yielding $x_{j}(n)$ from $y_{k}\left( n\right) $ (replacing of course the 
\textit{variables} $v_{j}\left( n\right) $ with their expressions in terms
of the \textit{initial data} $v_{k}\left( 0\right) $ as given just above,
see (\ref{vjnAppB})); and to then verify once more that, if the $18$\textit{%
\ a priori generic} parameters $a_{j\ell},~b_{j\ell},~c_{j\ell}$ do satisfy (%
\textit{only}!) the $2$ \textit{simple constraints }(\ref{2IsoConds}), then
the solution thus obtained does indeed satisfy the \textit{isochronicity}
conditions (\ref{xniso}).
\end{subequations}

\end{document}